\documentclass[pre,twocolumn,showpacs]{revtex4}
\usepackage{graphicx}
\usepackage[usenames]{color}
\newcommand{\beq}{\begin{equation}}
\newcommand{\eeq}{\end{equation}}
\newcommand{\beqa}{\begin{eqnarray}}
\newcommand{\eeqa}{\end{eqnarray}}
\newcommand{\ba}{\begin{array}}
\newcommand{\ea}{\end{array}} 
\begin{document} 

\title{One-dimensional superfluid Bose-Fermi mixture:
mixing, demixing and bright solitons} 
\author{Sadhan K. Adhikari$^{1}$
\footnote{Email: adhikari@ift.unesp.br; 
URL: www.ift.unesp.br/users/adhikari} 
and Luca Salasnich$^{2}$
\footnote{Email: salasnich@pd.infn.it; URL: www.padova.infm.it/salasnich}
}
\affiliation{
$^{1}$Instituto de F\'isica Te\'orica, UNESP $-$ S\~ao Paulo
State University,
01.405-900 Sao Paulo, Sao Paulo, Brazil 
\\
$^2$CNISM and CNR-INFM, Unit\`a di Padova, 
Dipartimento di Fisica ``G. Galilei'', Universit\`a di Padova, 
Via Marzolo 8, 35131 Padova, Italy } 

\begin{abstract} 
We study 
a ultra-cold and dilute   superfluid Bose-Fermi mixture
confined in a strictly one-dimensional atomic 
waveguide 
by 
using  a set of coupled 
nonlinear mean-field equations obtained 
from  the Lieb-Liniger energy density for bosons 
and the Gaudin-Yang energy density for fermions. 
We consider a finite Bose-Fermi inter-atomic strength $g_{bf}$ 
and both periodic and open boundary conditions. 
We find that with periodic boundary conditions, i.e. in a quasi-1D ring, 
a uniform Bose-Fermi mixture is stable only with a large fermionic density. 
We predict that at small fermionic densities 
the ground state of the system displays demixing if $g_{bf}>0$ 
and may become a localized Bose-Fermi bright soliton for $g_{bf}<0$. 
Finally, we show, using variational and numerical solution of the 
mean-field equations, that with open boundary conditions, 
i.e. in a quasi-1D cylinder, 
the Bose-Fermi bright soliton is the unique ground state of the system 
with a finite number of particles, which could exhibit a partial
mixing-demixing transition. In this case the bright solitons are 
demonstrated to be dynamically stable. The experimental  realization of 
these Bose-Fermi bright solitons seems possible with present setups. 
\end{abstract}


\maketitle

\section{Introduction}

Effects of quantum statistics are strongly enhanced in strictly 
one-dimensional (1D) systems \cite{takahashi-book}. 
These effects can be investigated with 
ultra-cold and dilute gases of alkali-metal atoms, which 
are now actively studied in the regime 
of deep Bose and Fermi degeneracy \cite{leggett-book}. 
Recently two experimental groups \cite{tg-psu,tg-mpi} have reported
the observation of the crossover between a 1D 
quasi Bose-Einstein condensate (BEC) in the weak-coupling 
mean-field Gross-Pitaevskii (GP) domain
and a Tonks-Girardeau (TG) gas \cite{girardeau,ll} 
with ultra-cold $^{87}$Rb atoms 
in highly elongated traps. A rigorous theoretical 
analysis of the ground-state 
properties of a uniform 1D Bose gas was performed by Lieb and Liniger (LL)
forty four years ago \cite{ll}. An extension of 
the LL theory for finite and inhomogeneous 
1D Bose gases has been
proposed on the basis the local density approximation (LDA) \cite{lda}. 
The LDA is improved by including a gradient term that represents 
additional kinetic energy associated with the inhomogeneity 
of the gas \cite{os,lsy,sala-ll2,sala-npse,sala-reduce,sala-ll}.  

In the last few years several
experimental groups have observed 
the crossover from the Bardeen-Cooper-Schrieffer (BCS)
state of Cooper Fermi pairs to the BEC 
of molecular dimers with ultra-cold two-hyperfine-component Fermi
vapors of $^{40}$K atoms \cite{greiner,regal,kinast}
and $^6$Li atoms \cite{zwierlein,chin}. 
It is well known that purely attractive potentials have bound states
in 1D and 2D for any strength, contrary to the 3D case \cite{landau}. 
A rigorous theoretical analysis of the ground-state
properties of a uniform 1D Fermi gas 
was performed by Gaudin and Yang (GY) 
forty years ago \cite{gy}. For repulsive interaction 
the GY model gives a Tomonaga-Luttinger liquid \cite{luttinger}, 
while for attractive interaction 
it describes a Luther-Emery (superfluid) liquid \cite{luther-emery}. 
The ground state of a weakly attractive 1D Fermi gas 
is a BCS-like state with Cooper pairs, whose size is much 
larger than the average inter-particle spacing. The 
strong coupling regime with tightly bound dimers 
is reached by increasing the magnitude of the 
attractive inter-atomic strength. In this regime the 
fermion pairs behave like a hard core Bose gas 
(TG gas), or equivalently like 1D noninteracting 
fermions \cite{recati,recati2}.  

Degenerate Bose-Fermi mixtures of alkali-metal atoms  
have been experimentally observed 
in $^{6,7}$Li \cite{exp3,mix-ex}, $^{6}$Li-$^{23}$Na \cite{exp4}, and 
$^{40}$K-$^{87}$Rb \cite{exp5}. 
In these mixtures, the theoretical investigation of phase separation 
\cite{molmer,nygaard,stoof,pethick,viverit,das,sala-toigo} 
and soliton-like structures has drawn significant attention. 
Bright solitons have been observed in BEC's of Li \cite{Li-soliton} and 
Rb \cite{Rb-soliton} atoms and studied subsequently 
\cite{BECsolitons}. It has been demonstrated using  microscopic 
\cite{BFsoliton(BBrepBFattr)} and mean-field hydrodynamic 
\cite{Sadhan-BFsoliton} models 
that the formation of stable fermionic bright and dark solitons is 
possible in a
degenerate Bose-Fermi mixture  as well as in a
Fermi-Fermi mixture \cite{ff} with the fermions in the normal state
in the presence of a sufficiently
attractive interspecies interaction which can overcome the Pauli
repulsion among fermions. The formation of a soliton in these cases is
related to the fact that the system can lower its energy by forming high
density regions (solitons) when the interspecies attraction is large
enough to overcome the Pauli repulsion in the degenerate Fermi gas  (and 
any possible repulsion in the BEC) \cite{skbs}. 
There have also been studies of mixing-demixing transition 
in degenerate Bose-Fermi \cite{sala-sadhan2}
and Fermi-Fermi \cite{skabm} mixtures with fermions in the normal state. 

After observing the degenerate Fermi gas and the realization of BCS 
condensed superfluid phase of the 
fermionic system \cite{bcsexp}, the BCS-Bose crossover \cite{cross} 
in it seems to be under control \cite{greiner,regal,kinast,zwierlein,chin}
by manipulating the Fermi-Fermi interaction near a Feshbach 
resonance \cite{fesh} by varying a uniform  external magnetic field.  
Naturally, the question of the possibility of bright solitons in a 
superfluid fermionic condensate assisted 
by a BEC (with attractive interspecies 
Bose-Fermi interaction) needs to be revisited. Although the fermionic 
BCS or molecular dimer  phase is possible with an attractive 
Fermi-Fermi attraction, once formed such a phase
is inherently repulsive and self-defocussing and will not lead to 
a bright soliton spontaneously. 

In this paper we investigate a 1D superfluid Bose-Fermi mixture composed 
of  bosonic atoms, 
well described by the LL theory, and superfluid fermionic atoms, 
well described by the GY theory with the special intention of studying 
the localized structures or bright solitons in this superfluid mixture. 
We derive a set of coupled nonlinear mean-field equations for the 
Bose-Fermi mixture which we use in the present study.
The solution of this mean-field equation 
is  considered  two types of boundary conditions: periodic and open. 
It has been recently shown 
\cite{ueda1,ueda2,sala-ring} that an attractive BEC 
with periodic boundary conditions, which can be experimentally 
produced with a quasi-1D ring \cite{gupta}, 
displays a quantum phase transition from 
a uniform state to a symmetry-breaking 
state characterized by a localized bright-soliton condensate 
\cite{sala-ring}. Here we show that a similar phenomenon 
appears in the 1D Bose-Fermi mixture with an attractive Bose-Fermi 
scattering length. Instead, with a repulsive Bose-Fermi 
scattering length or with 
the increase of the number of Fermi atoms leading to large repulsion    
we find a phase separation between bosons and 
fermions (a mixing-demixing transition), in analogy with previous 
theoretical 
\cite{molmer,nygaard,stoof,pethick,viverit,das,sala-toigo} 
and experimental studies \cite{mix-ex} with 3D trapped $^6$Li-$^7$Li 
and $^{39}$K-$^{40}$K mixtures. Finally, we predict that with open boundary 
conditions, i.e. in a infinite quasi-1D cylinder, 
the ground-state of the mixture 
with sufficiently attractive Bose-Fermi interaction is always a 
localized bright soliton. 

The paper is organized as follows. 
In Sec. II we describe the 1D model we use in our 
investigation of the superfluid Bose-Fermi mixture. The Lagrangian 
density for bosons is appropriate for a TG to BEC crossover through the 
use of a quasi-analytic LL function. The fermions are treated 
using the GY model through a quasi-analytic GY function. 
The Bose-Fermi interaction is taken to be a standard contact interaction. The 
Euler-Lagrange equations for this system are a set of coupled nonlinear 
mean-field equations which we use in the present study. 
In Sec. III we consider the system with periodic boundary conditions 
and obtain the thresholds for demixing, for the formation of localized 
bright solitons (for a sufficiently strong 
Bose-Fermi attraction), and for the existence of mixing, i.e.  
states with constant density in space 
(for weak Bose-Fermi attraction and for Bose-Fermi repulsion). 
We also present a modulational instability analysis of a uniform 
solution of the mixture and obtain the condition 
for the appearance of bright solitons by
modulational instability. 
In Sec. IV we consider the mixture with open boundary conditions  
and derive a variational approximation for the solution 
of the mean-field equations of the model using a Gaussian-type ansatz. 
We investigate numerically the bright Bose-Fermi 
solitons, demonstrate their stability under perturbation, 
and compare the numerical results  with those of the   variational 
approach. 
Finally, in Sec. V we present a summary and discussion of our study. 

\section{The model} 

We consider a mixture of $N_b$ atomic 
bosons of mass $m_b$ and $N_f$ superfluid  
atomic fermions of mass $m_f$ at zero temperature 
trapped by a tight cylindrically symmetric harmonic 
potential of frequency $\omega_{\bot}$ in the 
transverse (radial cylindric) direction. We assume factorization 
of the transverse degrees of freedom. 
This is justified in 1D confinement where, regardless
of the longitudinal behavior
or statistics, the transverse spatial profile is that
of the single-particle ground-state \cite{das,sala-st}. 
The transverse width of the atom distribution  
is given by the characteristic harmonic length of the 
single-particle ground-state: 
$a_{\bot j}=\sqrt{\hbar/(m_j\omega_{\bot})}$, with $j=b,f$. 
The atoms have an effective 1D 
behavior at zero temperature if their 
chemical potentials are much smaller than 
the transverse energy $\hbar\omega_{\bot}$ \cite{das,sala-st}. 
The boson-boson interaction is described 
by a contact pseudo potential with repulsive (positive) 
scattering length $a_b$. 
The fermion-fermion interaction is modelled by 
a contact pseudo potential with attractive (negative) 
scattering length $a_f$. 
The boson-fermion interaction is instead characterized 
by a contact pseudo potential with 
scattering length $a_{bf}$, which can be repulsive 
or attractive. To avoid the confinement-induced resonance 
\cite{olshanii} we take $a_b,|a_f|,|a_{bf}|\ll a_{\bot b},a_{\bot f}$. 

We use a mean-field effective Lagrangian 
to study the static and collective properties 
of the 1D Bose-Fermi mixture. 
The Lagrangian density ${\cal L}$ of the mixture reads 
\beq 
{\cal L} = {\cal L}_b + {\cal L}_f + {\cal L}_{bf} \; . 
\label{l-tot}
\eeq 
The term ${\cal L}_b$ is the bosonic Lagrangian, defined as 
\beqa
{\cal L}_b  = i\hbar\, \psi_b^* {\partial \over \partial t} \psi_b 
+ {\hbar^2\over 2m_b} \psi_b^* {\partial^2 \over \partial z^2} \psi_b 
\nonumber
\\
- {\hbar^2\over 2m_b} 
|\psi_b|^6 G\left({m_b g_b \over \hbar^2 |\psi_b|^2}\right) \; , 
\label{l-b}
\eeqa
where $\psi_b(z,t)$ is the hydrodynamic field of the Bose gas
along the longitudinal axis, 
such that $n_b(z,t)=|\psi_b(z,t)|^2$ is the 1D local probability 
density of bosonic atoms  and $g_b=2\hbar \omega_{\bot}\, a_b$ 
is the 1D boson-boson interaction strength ($g_b>0$).
 $G(x)$ is the Lieb-Liniger function \cite{pade-b}, 
defined as the solution of a Fredholm equation for $x>0$, 
and such that \cite{ll} 
\beq
G(x) =
\left\{
\begin{array}{cc}
x -{4\over 3\pi}x^{3/2}+ ...  
&
\mbox{ for } 0 < x \ll 1, 
\\
{\pi^2\over 3}(1 - {2\over x} + ... ) 
&
\mbox{ for } x \gg 1.
\\
\end{array}
\right.
\eeq
In the extreme weak-coupling limit $x \to +0$, and $G(x)\to x$ and 
consequently, the bosonic Lagrangian above reduces to the standard 
mean-field GP Lagrangian. 
In the static case the Lagrangian density ${\cal L}_b$ reduces exactly
to the energy functional recently introduced by
Lieb, Seiringer and Yngvason \cite{lsy}. 
In addition, ${\cal L}_b$ has been successfully
used to determine the collective oscillation
of the 1D Bose gas with longitudinal
harmonic confinement \cite{sala-ll}. In the strong coupling limit $x \to 
+\infty$, and $G(x)\to \pi^2/3$ while the Lagrangian above reduces to 
the strongly repulsive bosonic Lagrangian in 
the  TG limit \cite{girardeau}.  As $x$ changes from $+0$ to $+\infty$ 
the above Lagrangian shows a continuous transition from the GP BEC to TG 
phase. 

The fermionic Lagrangian density ${\cal L}_f$ is 
given instead by 
\beqa
{\cal L}_f  = i\hbar\, \psi_f^* {\partial \over \partial t} \psi_f 
+ {\hbar^2\over 2m_f} \psi_f^* {\partial^2 \over \partial z^2} \psi_f
\nonumber 
\\
- {\hbar^2\over 2m_f}
|\psi_f|^6 F\left({m_f g_f \over \hbar^2 |\psi_f|^2}\right) 
 \; ,
\label{l-f} 
\eeqa
where $\psi_f(z,t)$ is the field 
of the 1D superfluid Fermi gas, such that $n_f(z,t)=|\psi_f(z,t)|^2$ 
is the 1D fermionic local density along the longitudinal axis  and 
$g_f=2 \hbar\omega_{\bot}\, a_f$ is the 1D fermion-fermion
interaction strength ($g_f<0$). 
$F(x)$ is the Gaudin-Yang function \cite{pade-f}, 
defined as the solution of a Fredholm equation for $x<0$, 
and such that \cite{gy,recati} 
\beq 
\label{5}
F(x) = 
\left\{ 
\begin{array}{cc}
{\pi^2\over {48}} (1 - {1\over x} + {3\over 4 x^2} + ... 
) 
& 
\mbox{ for } x \ll -1, 
\\
{\pi^2\over {12}}(1 + {6\over \pi^2} x + ...) 
&
\mbox{ for } -1 \ll x < 0. 
\\
\end{array}
\right.
\eeq
In the static and uniform case the Lagrangian density ${\cal L}_f$ 
reduces exactly to Gaudin-Yang energy functional \cite{gy}, 
that has been recently used by Fuchs {\it et al.} \cite{recati}. 

{ 
The physical content of the fermionic Lagrangian (\ref{l-f}) is easy to 
understand. For 
$x\to -0$ we are in the domain of weak Fermi-Fermi 
attraction ($a_f, g_f \to 0$ corresponding to the BCS limit), while 
$F(x)\to \pi^2 /{12}$ from 
Eq. 
(\ref{5}). 
Consequently, 
the fermionic interaction term  in Eq. (\ref{l-f}) involving the 
$F(x)$ function reduces to $(\hbar^2/2m_f)(\pi^2/{12) 
|\psi_f|^6}$ (independent of Fermi-Fermi interaction strength $g_f$ or 
Fermi-Fermi scattering length $a_f$), 
which 
is the Pauli repulsive term of 
noninteracting fermions considered in 
Ref. \cite{sala-st,sala-fermi} arising due 
to the Pauli exclusion principle and not due to the fundamental 
Fermi-Fermi interaction implicit in the scattering length $a_f$ via 
$g_f$. This is quite 
expected as by taking the $x\to -0$ limit we switch off the Fermi-Fermi 
attraction and pass from the BCS domain to a degenerate noninteracting 
Fermi gas (in non-superfluid phase) studied in Ref. \cite{sala-st} 
corresponding to $x\to +0$. 
[Although the probability densities in the $x\to +0$ (non-superfluid 
degenerate Fermi gas)
and $x\to -0$  (superfluid BCS condensate)
limits are identical, they correspond to different physical states. The 
superfluid phase gas a ``gap" associated with it describable by the BCS 
equation.]
As $x$ passes  from $-0$ to 
$-\infty$, the Fermi-Fermi attraction increases and we move from the  
superfluid BCS regime of weakly-bound Cooper pairs to the unitarity 
regime of strongly-bound molecular  fermions. (The unitarity limit 
corresponds to infinitely large Fermi-Fermi attraction: $a_f \to 
-\infty$.) 
As Fermi-Fermi scattering length
$|a_f|$ increases from the BCS to the unitarity limit,  the interaction 
energy in Eq. (\ref{l-f}) gets a term 
dependent on $a_f$. 
However,  in the unitarity limit $a_f \to -\infty$ and 
all the 
fermions will be paired to form strongly bound noninteracting molecules 
and the Fermi-Fermi interaction term in Eq. (\ref{l-f}) behaves like 
that of 
a 
Tonks-Girardeau gas of bosons $-$ bosonic molecules  $-$ and again 
becomes independent of  
$a_f$ 
\cite{recati}.  The system then becomes a Bose gas of molecules and its 
interaction energy is no longer a function of $a_f$.
 }

Finally, the Lagrangian density ${\cal L}_{bf}$ 
of the Bose-Fermi interaction taken to be of the following standard 
zero-range form \cite{sala-st,sala-sadhan2}
\beq 
{\cal L}_{bf} = - g_{bf} \, 
|\psi_b|^2 |\psi_f|^2 \; ,  
\label{l-bf} 
\eeq 
where $g_{bf}=2 \hbar\omega_{\bot} a_{bf}$ 
is the 1D boson-fermion interaction strength. 

{
The Euler-Lagrange equations of the Lagrangian
${\cal L}$ are the two following
coupled partial differential equations:
\beqa
i\hbar \partial_t \psi_b &=& \biggr[ -\frac{\hbar^2}{2m_b} \partial_z^2
+ 3\frac{\hbar^2}{2m_b} n_b^2 G\biggr({m_b g_b\over \hbar^2n_b}\biggr) 
\nonumber 
\\
&-& {1\over 2}g_bn_b
G'\biggr({m_bg_b\over \hbar^2n_b}\biggr)
 + g_{bf} n_f \biggr] \psi_b  \; ,
\label{str1a}\\
i \hbar \partial_t \psi_f &=& \biggr[ - \frac{\hbar^2}{2m_f} 
\partial_z^2
+ 3\frac{\hbar^2}{2m_f} 
n_f^2 F\biggr({m_fg_f\over \hbar^2 n_f}\biggr) \nonumber \\
&-& {1\over 2}g_fn_f
F'\biggr({m_f g_f\over \hbar^2 n_f}\biggr)
 + g_{bf} n_b \biggr] \psi_f  \; ,
\label{str2a}
\eeqa
where $n_j=|\psi_j|^2, j=b,f$ 
are probability densities for bosons and 
fermions, respectively, 
with the normalization $\int_{-\infty}^\infty n_j dz = N_j$.}

{It is convenient to work in terms of dimensionless 
variables defined in 
terms of a frequency $\omega$ and length  
$l\equiv \sqrt{\hbar / (m_b \omega)}$ by 
$\psi_j=\hat \psi_j/\sqrt l$, 
$t= 2 \hat t/\omega$,  $z=\hat z l$, $g_j={\hat g_j \hbar^2/(2 m_bl)}$,   
and $g_{bf}={\hat g_{bf} \hbar^2/(2 m_bl)}$. In terms of these new 
variables Eqs. (\ref{str1a}) and (\ref{str2a})
can be written as
\beqa
i \partial_t \psi_b &=& \biggr[ - \partial_z^2
+ 3n_b^2 G\biggr({g_b\over 2n_b}\biggr) \nonumber \\
&-& {1\over 2}g_bn_b
G'\biggr({g_b\over 2n_b}\biggr)
 + g_{bf} n_f \biggr] \psi_b  \; ,
\label{str1}\\
i \partial_t \psi_f &=& \biggr[ - \lambda  \partial_z^2
+ 3\lambda  n_f^2 F\biggr({g_f\over 2n_f}\biggr) \nonumber \\
&-& {1\over 2}g_fn_f
F'\biggr({g_f\over 2n_f}\biggr)
 + g_{bf} n_b \biggr] \psi_f  \; ,
\label{str2}
\eeqa
where we have dropped the hats over the variables, and where  
$\lambda= m_b/m_f,$  
$n_j=|\psi_j|^2, j=b,f$ 
are  probability densities for bosons and 
fermions, respectively, 
with the normalization $\int_{-\infty}^\infty n_j dz = N_j$. However, 
in the present study we  
set $\lambda =1$ in the following. All results reported in this paper 
are for this case. This should approximate well the Bose-Fermi mixtures 
$^7$Li-$^6$Li and $^{39}$K-$^{40}$K of experimental interest.}

The bosonic and fermionic nonlinearities in Eqs. (\ref{str1}) and 
(\ref{str2}), respectively,  have complex 
structures in general. However, in the weak-coupling GP limit  ($g_b \to 
0$) the bosonic nonlinearity in Eq. (\ref{str1}) is cubic and turns 
quintic in the strong-coupling TG limit ($g_b \to
+\infty$).  The fermionic nonlinearity  in Eq. (\ref{str2})
becomes quintic in both weak ($g_F\to -0$)
and strong ($g_f \to
-\infty$)
coupling but with coefficients $\pi^2/{12}$ and 
$\pi^2/{48}$, respectively.   

For stationary states the solution of Eqs. (\ref{str1}) and (\ref{str2})
have the form $\psi_j=\phi_j \exp(-i\mu_jt)$ where $\mu_j$ are the 
respective chemical potentials. Consequently, these equations reduce to  
\beqa
\mu_b \phi_b &=& \biggr[ - \partial_z^2
+ 3n_b^2 G\biggr({g_b\over 2n_b}\biggr) \nonumber \\
&-& {1\over 2}g_bn_b
G'\biggr({g_b\over 2n_b}\biggr)
 + g_{bf} n_f \biggr] \phi_b  \; ,
\label{mu1}\\
\mu_f \phi_f &=& \biggr[ - \partial_z^2
+ 3n_f^2 F\biggr({g_f\over 2n_f}\biggr) \nonumber \\
&-& {1\over 2}g_fn_f
F'\biggr({g_f\over 2n_f}\biggr)
+ g_{bf} n_b \biggr] \phi_f  \; .
\label{mu2}
\eeqa
A repulsive Bose-Bose interaction is produced by a positive 
$g_b$, while an attractive Fermi-Fermi interaction 
corresponds to a negative $g_f$.  

\section{Mixture with periodic boundary conditions} 

\subsection{General Considerations}

Here we consider the 1D Bose-Fermi mixture with periodic 
boundary 
conditions. These boundary conditions can be used to model 
a quasi-1D ring of radius $R$. 
If the radius is much larger than the transverse width, 
i.e. $R \gg a_{\bot j}$, $j=b,f$, then effects of curvature 
can be neglected \cite{sala-ring} and one can safely 
use the Lagrangian density (\ref{l-tot}) with $z=R\theta$, 
where $\theta$ the azimuthal angle \cite{sala-ring,gupta}. 
The energy density of the uniform mixture 
is immediately derived from the Lagrangian 
density (\ref{l-tot}), with (\ref{l-b}), (\ref{l-f}) and (\ref{l-bf}), 
dropping out space-time derivatives. It is given by 
\beq
{\cal E} = n_b^3 G\left({g_b \over 2 n_b}\right) + 
n_f^3 F\left({g_f \over 2 n_f}\right) + g_{bf} \, n_b \, n_f \; .   
\label{energy-den}
\eeq 
In this case one can have a uniform mixture or the formation of  
soliton-like structures
and in the following we study the condition for these possibilities 
to happen. 
Obviously, for the 1D uniform mixture in the ring of radius $R$, 
we have  $n_b=N_b/(2\pi 
R)$ and $n_f=N_f/(2\pi R)$. 

\begin{figure}[tbp]
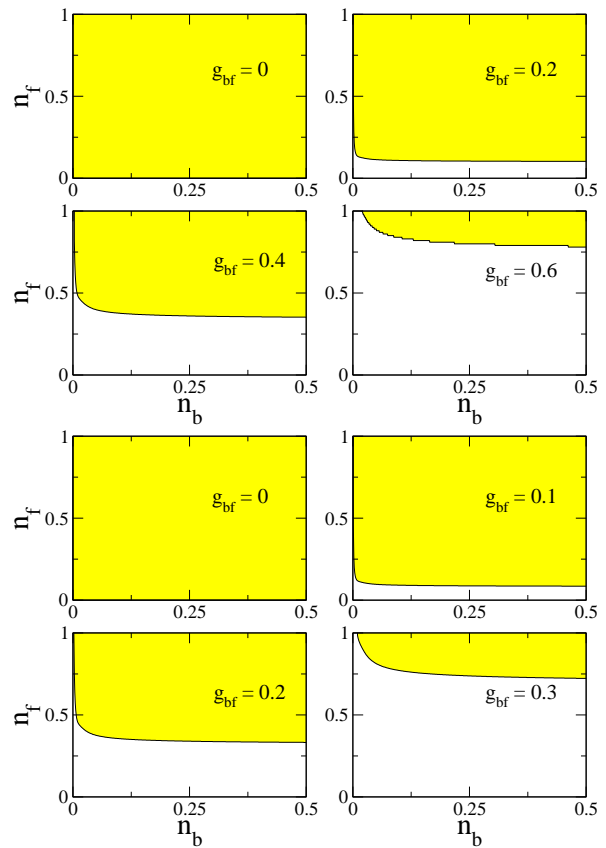

\begin{center}
{\includegraphics[width=.9\linewidth,clip]{fig1a_new.eps}}
{\includegraphics[width=.9\linewidth,clip]{fig1b_new.eps}}
\end{center}
\caption{(Color online) Region of 
stability 
of the homogeneous mixture denoted by shaded (gray) area
in the plane $(n_b,n_f)$. 
Different panels correspond to different values of the 
Bose-Fermi interaction strength $g_{bf}$. We set $g_b=0.1$ 
and (a) $g_f=-0.1$ (corresponding to a weak Fermi-Fermi attraction in 
the BCS phase) and (b) $g_f=-30$ (corresponding to a strong Fermi-Fermi 
attraction, with fermions in the unitarity regime).}
\label{fig1}
\end{figure}

The uniformly  mixed phase is energetically stable if its energy is a 
minimum 
with respect to small variations of the densities $n_f$ and $n_b$, while the 
total number of fermions and bosons are held fixed. 
To get the equilibrium densities one must then minimize the function
\beq
{
{\tilde {\cal E}}(n_b,n_f) = 
{\cal E}(n_b,n_f) - \mu_b(n_b^*,n_f^*) n_b - \mu_f(n_b^*, n_f^*) n_f} 
\eeq
where $\mu_b$ and $\mu_f$ are Lagrange multipliers (imposing that the
numbers of bosons and fermions are fixed) which may be identified with the
Bose and Fermi chemical potentials {
and $n_b^*$ and  $n_f^*$ are the 
values of $n_b$ and  $n_f$ at the minimum. }
Setting the derivatives of $\tilde{{\cal E}}$ 
with respect to $n_f$ and $n_b$ equal to zero, one finds:
\beq
\mu_b = {\partial {\cal E}\over \partial n_b} = 
3n_b^2 G\biggr({g_b\over 2n_b}\biggr) 
- {1\over 2}g_b n_b G'\biggr({g_b\over 2n_b^2}\biggr)
+ g_{bf} n_f \; , 
\label{chem-b}
\eeq
\beq
\mu_f = {\partial {\cal E}\over \partial n_f} = 
3n_f^2 F\biggr({g_f\over 2n_f}\biggr)
- {1\over 2}g_f n_f F'\biggr({g_f\over 2n_f^2}\biggr) 
+ g_{bf} n_b \; .  
\label{chem-f} 
\eeq 
The solution of Eqs. (\ref{chem-f}) and (\ref{chem-b}), 
which are exactly Eqs. (\ref{mu1}) and (\ref{mu2}) if one 
drops out the space derivatives, 
gives a minimum if the corresponding Hessian
of $\tilde{{\cal E}}$ is positive, i.e. if: 
\beq
{\frac{\partial^2 \tilde{{\cal E}} }{\partial n_b^2}} {\frac{\partial^2
\tilde{{\cal E}} }{\partial n_f^2}} - 
\left( {\frac{\partial^2 \tilde{{\cal E}} }
{\partial n_b \partial n_f}} \right)^2 > 0 \; , 
\label{ineq} 
\eeq 
The solution of this inequality gives the region 
in the parameters' space where the homogeneous mixed
phase is energetically  stable.

In Fig. \ref{fig1} we show  the region of  
stability of the homogeneous mixture by the shaded (gray) area 
in the plane $(n_b,n_f)$ 
for different values of $g_{bf}$. Note that the sign of $g_{bf}$ 
is not important because in Eq. (\ref{ineq}) appears $g_{bf}^2$. 
The figure shows that, by increasing $g_{bf}$, 
the uniform mixture is stable only at large 
fermionic densities $n_f$. This result, that is in agreement 
with previous 1D predictions \cite{das} on a 1D mixture 
of Bose-condensed atoms and normal Fermi atoms,   
is exactly the opposite of what happens in a 3D mixture 
of bosons and fermions. In fact, a uniform 3D mixture 
is stable only for small values of the fermionic density 
\cite{molmer,nygaard,stoof,pethick,viverit,sala-toigo}. 
Figure  1 also shows that when bosons enter in the TG regime 
(i.e. $g_b/n_b\gg 1$) the uniform mixture 
is much less stable: see the behavior of the stability 
line at very small $n_b$, with $g_{bf}\ne 0$. 

\begin{figure}[tbp]
\begin{center}
{\includegraphics[width=\linewidth,clip]{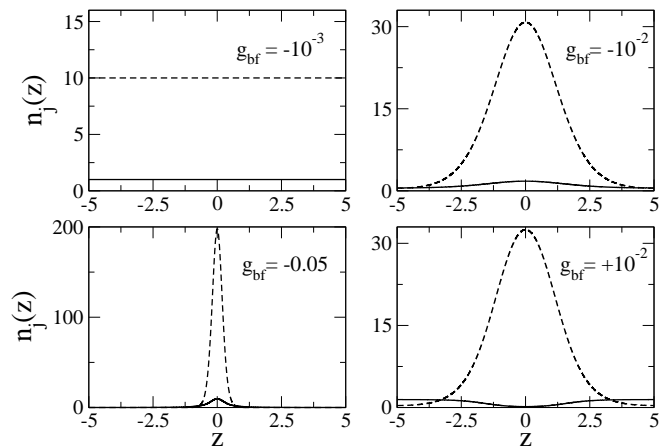}}
\end{center}
\caption{Probability density profiles $n_b(z)$ of bosons (dashed line)
and $n_f(z)$ of fermions (solid line)
of the 
Bose-Fermi mixture 
with $N_b=100$ bosons and $N_f=10$ fermions and periodic 
boundary conditions (axial length $L\equiv 2 \pi R=10$). We choose 
$g_b=0.01$ and $g_f={-0.025}$ and calculate the profiles 
for different values of the Bose-Fermi interaction 
strength $g_{bf}$.} 
\label{fig-figata}
\end{figure}

The panels of Fig. 1 suggest that, 
with a finite $g_{bf}$, at small fermionic densities 
the uniform mixture is unstable: 
the ground-state of the system displays demixing if $g_{bf}>0$ 
and becomes a localized Bose-Fermi bright soliton if $g_{bf}<0$. 
These effects are clearly shown in Fig. \ref{fig-figata}, 
where we plot the density profiles $n_j(z)$, with $j=b,f$, 
for different values of $g_{bf}$ 
calculated by directly solving Eqs. (\ref{mu1}) and (\ref{mu2}) 
numerically with 
appropriate boundary conditions. 

The profiles of probability densities plotted in Fig. \ref{fig-figata} 
demonstrate the uniform-to-localized and the uniform-to-demixed 
transition. In fact, a large and negative $g_{bf}$ induces a strong 
localization while a large and positive $g_{bf}$ produces 
demixing. These density profiles have been obtained 
by solving Eqs. (\ref{str1}) and (\ref{str2}) with 
a finite-difference Crank-Nicholson algorithm and 
imaginary time \cite{sala-crank}. 

In the regime $g_b/n_b\ll 1$ (BEC limit)
and $|g_f|/n_f \ll 1$ (BCS limit), the 
above analysis yields simple analytic results. In this case 
${\cal E}= g_b n_b^2/2+\pi^2n_f^3/{12}+g_{bf}n_bn_f$, 
consequently, Eq.  (\ref{ineq}) leads
to the condition 
\begin{equation}\label{con1}
{\frac{1}{2}}
\pi^2g_b n_f >g_{bf}^2
\end{equation}
for the stability of the uniform mixture. 
The condition $\pi^2g_b n_f/{2} >g_{bf}^2$ is 
qualitatively 
consistent with Fig. \ref{fig1} (a). For example, 
for $g_b= 0.1$ and  $g_f=-0.1$ the condition of stability of uniform mixture 
becomes $n_f > {2} g_{bf}^2 $ and 
for \textcolor{Red}{
$|g_{bf}| = 0, 0.2,{0.4,}$ and {0.6} leads to $n_f> 
0,
{ 0.08, 0.32,}$ and {0.72} respectively. }

In the regime $g_b/n_b\ll 1$ (BEC limit)
and $|g_f/n_f|\gg 1$ (unitarity or tightly
bound molecule formation limit) 
${\cal E}= g_b
n_b^2/2+\pi^2n_f^3/{48}+g_{bf}n_bn_f$, consequently, Eq.  
(\ref{ineq})
leads to the condition \begin{equation} \label{con2}{ 
{1\over 8}} \pi^2g_b
n_f >g_{bf}^2. \end{equation} for the stability of the uniform mixture.
For example, for $g_b= 0.1$ and $g_f=-30$ the condition of stability of
uniform mixture becomes $n_f > {8} g_{bf}^2 $ and for 
\textcolor{Red}{$|g_{bf}| = 0,
0.1, {0.2},$ and {0.3} leads to $n_f> 0, 
{0.08, 0.32,}$ and {0.72},} respectively, in
qualitative agreement with Fig. \ref{fig1} (b). The time-independent
conditions (\ref{con1}) and (\ref{con2}) are also derived in following
subsection using rigorous time-dependent modulational instability
analysis of the uniform mixture (constant-amplitude solutions) in the
weak and strong Fermi-Fermi coupling limits.

Finally, in the regime $g_b/n_b\gg 1$ (TG limit)
and $|g_f/n_f|\ll 1$ (BCS limit) 
${\cal E}= \pi^2
n_b^3/3+\pi^2n_f^3/{12}+g_{bf}n_bn_f$, 
consequently, Eq.  (\ref{ineq})
leads to the condition \begin{equation} \label{con6}  \pi^4  n_b
n_f >g_{bf}^2. \end{equation} for the stability of the uniform mixture.
The functional dependence of 
condition (\ref{con6}) on $n_b,n_f, g_b$  and $g_f$ is qualitatively 
different from conditions 
(\ref{con1}) and (\ref{con2}).

The inequality (\ref{ineq}) can be written as 
\beq
c_b^2 c_f^2 > 4 g_{bf}^2 n_b n_f 
\; ,  
\label{ineq2} 
\eeq
where 
$c_f=\sqrt{2 n_f (\partial \mu_f/\partial n_f)}$ 
is the sound velocity of the superfluid Fermi component 
and $c_b=\sqrt{2 n_b (\partial \mu_b/\partial n_b)}$
is the sound velocity of the Bose gas.
The sound velocity $c_{bf}$ of the 1D Fermi-Bose mixture can be 
easily obtained following a procedure similar to the one suggested 
by Alexandrov and Kabanov \cite{kabanov} 
for a two-component BEC. One finds:
\beq
c_{bf}= {\frac{1}{\sqrt{2}}} \sqrt{c_b^2+c_f^2 \pm 
\sqrt{(c_b^2-c_f^2)^2 + 16 g_{bf}^2 n_b n_f }} \; . 
\eeq
Thus the sound velocity has two branches and the homogeneous mixture 
becomes dynamically unstable when the lower branch becomes imaginary. 

\subsection{Modulational Instability}

We show in the following that for attractive Bose-Fermi interaction,
the transition from uniform mixture to localized soliton-like structures 
considered in Sec. IIIA is due to modulational instability.  
To study analytically the modulational instability 
\cite{sala-prl,sala-nick} 
of Eqs. (\ref{str1}) and (\ref{str2}) 
we consider the special 
case of weak Bose-Bose (small positive $g_b$)   and both weak (small 
$|g_f/n_f|\ll 1$ corresponding to BCS limit) and strong (large
$|g_f/n_f|\gg 1$ corresponding to molecular unitarity limit)  
Fermi-Fermi interactions  while these equations reduce to  
\beqa
i \partial_t \psi_b = \left[ - \partial_z^2
+ g_b|\psi_b|^2   - g_{bf} |\psi_f|^2 \right] \psi_b  \; ,
\label{m1}
\eeqa
\beqa
i \partial_t \psi_f = \left[ - \partial_z^2
+\kappa \pi^2 |\psi_f|^4
 - g_{bf} |\psi_b|^2 \right] \psi_f  \; ,
\label{m2}
\eeqa
where $\kappa=1/{4}$ for $|g_f/n_f|\ll 1$ and 
1/{12} for $|g_f/n_f|\gg 1$. Here   
we have taken the interspecies interaction to be attractive 
by inserting an explicit negative sign in  $g_{bf}$.

We analyze the modulational instability of a 
constant-amplitude solution corresponding to a uniform mixture 
in coupled
equations (\ref{m1}) and (\ref{m2}) by considering 
the solutions
\begin{eqnarray}
\varphi_{b0}=A_{b0}\exp(i\delta_b)\equiv A_{b0}e^{i t (g_{bf}
A_{f0}^2
-g_b  A_{b0}^2 )}, \label{s1}
\end{eqnarray}
\begin{eqnarray}
\varphi_{f0}=A_{f0}\exp(i\delta_f)\equiv A_{f0}e^{i t (g_{bf}
A_{b0}^2
-\kappa \pi^2  A_{f0}^{4} )}, \label{s2}
\end{eqnarray}
of  Eqs.  (\ref{m1}) and (\ref{m2}), respectively, where $A_{j0}$ is the 
amplitude and 
$\delta_j$ a phase
for component $j=b,f$. The constant-amplitude solution develops an
amplitude dependent phase on time evolution.
We consider a  small perturbation $A_j\exp(i\delta_j)$ to
these solutions via
\begin{eqnarray}
\varphi_j=(A_{j0}+A_j)\exp(i\delta_j),
  \end{eqnarray}
where $A_j=A_j(z,t)$.
Substituting these perturbed solutions in Eqs.  (\ref{m1}) and  
(\ref{m2}),
and for small perturbations retaining
only the linear terms in $A_j$ we get
\begin{eqnarray}\label{q1}
&i&{\partial_t  A_b}
+ {\partial_z ^2 A_b} 
-   g_b A_{b0}^{2}(A_b+A_b^*)\nonumber \\
&+&
g_{bf}
A_{b0}A_{f0} (A_f+A_f^*)=0, \\
\label{q2}
&i&{\partial_t A_f} + {\partial_z ^2 A_f} 
-   2\kappa \pi^2 A_{f0}^{4}(A_f+A_f^*)\nonumber \\
&+&g_{bf}
A_{b0}A_{f0} (A_f+A_f^*)=0. 
\end{eqnarray}
We consider the complex plane-wave
perturbation
\begin{equation}
A_j(z,t)=
{\cal A}_{j1} \cos (Kt -\Omega z)+i {\cal A}_{j2} \sin  (Kt 
-\Omega
z) \label{comp}
\end{equation}
with $j=b,f$, where ${\cal A}_{j1}$ and  
${\cal A}_{j2}$
are the amplitudes for the real and imaginary parts, respectively, and
$K$ and $\Omega$ are frequency and wave numbers.

Substituting  Eq. (\ref{comp}) in Eqs. (\ref{q1}) and (\ref{q2}) and  
separating the real and imaginary parts we get
\begin{eqnarray}\label{p1}
-{\cal A}_{b1}K&=&{\cal A}_{b2}\Omega^2, \\
-{\cal
A}_{b2}K&=&{\cal
A}_{b1}\Omega^2-2g_{bf}A_{b0}A_{f0}{\cal A}_{f1}+2g_b
A_{b0}^{2}{\cal A}_{b1}, \nonumber \\ \label{p2}
  \end{eqnarray}
for  $j=b$, and
\begin{eqnarray}\label{p3}
-{\cal   A}_{f1}K&=&{\cal   A}_{f2}\Omega^2, \\
-{\cal
A}_{f2}K&=&{\cal
  A}_{f1}\Omega^2-2g_{bf}A_{b0}A_{f0}{\cal A}_{b1}+4\kappa \pi^2
A_{f0}^{4}{\cal   A}_{f1},   \nonumber \\  \label{p4}
 \end{eqnarray}
for $j=f$.
Eliminating ${\cal A}_{b2}$ between Eqs.  (\ref{p1})  and
(\ref{p2}) we get 
\begin{eqnarray}\label{p5}
{\cal   A}_{b1}[K^2-\Omega^2( \Omega^2+2g_b   A_{b0}^{2})]=-
2{\cal A}_{f1}g_{bf}A_{b0}A_{f0} \Omega^2,  \end{eqnarray}
and eliminating ${\cal A}_{f2}$ between  (\ref{p3})  and
(\ref{p4}) we have 
\begin{eqnarray}\label{p6}
{\cal A}_{f1}[K^2-\Omega^2( \Omega^2+4\kappa \pi^2A_{f0}^{4})]=-
2{\cal A}_{b1}g_{bf}A_{b0}A_{f0} \Omega^2.  \end{eqnarray}
Finally, eliminating ${\cal A}_{b1}$ and ${\cal A}_{f1}$
from   (\ref{p5}) and (\ref{p6}), we obtain the following
dispersion relation
\begin{eqnarray}\label{p7}
&K&= \pm \Omega\biggr[ \left(\Omega^2+
g_{b}A_{b0}^{2}
+
2\kappa \pi^2A_{f0}^{4}\right)\nonumber \\ 
&\pm& \left\{ \left(
g_{b}A_{b0}^{2} -
2\kappa \pi^2A_{f0}^{4} \right)^2
+
4g_{bf}^2
A_{b0}^2A_{f0}^2 \right\}^{1/2} \biggr]^{1/2}.
\end{eqnarray}

For stability of the plane-wave perturbation,  $K$ has to be
real. For any $\Omega$ this happens for
\begin{eqnarray}
(g_{b}A_{b0}^{2}
+2\kappa \pi^2 A_{f0}^{4})^2 &>& \left(
g_{b}A_{b0}^{2} -
2\kappa \pi^2 A_{f0}^{4} \right)^2 \nonumber \\ &+&
4g_{bf}^2
A_{b0}^2A_{f0}^2, 
\end{eqnarray}
or for 
\begin{equation}\label{con3}
2g_b \kappa \pi^2 A_{f0}^2 > g_{bf}^2.
\end{equation}
However, for
$2g_b\kappa  \pi^2 A_{f0}^2 < g_{bf}^2$, 
$K$ can become imaginary and the plane-wave perturbation 
can grow exponentially with time. This is the domain of modulational
instability of a constant-amplitude solution (uniform mixture of Sec. 
IIIA)
signalling the possibility
of coupled Bose-Fermi bright soliton to appear.
Noting that $A_{f0}^2=n_f$ for the uniform mixture, 
this result is consistent with the stability analysis of 
Sec. IIIA. In the weak-coupling BCS limit $\kappa=1/{4}$ 
and Eq. 
(\ref{con3}) 
reduces to Eq. (\ref{con1}) obtained from energetic considerations. 
In the strong coupling  unitarity limit  $\kappa=1/{12}$
and  Eq. (\ref{con3}) reduces to Eq. (\ref{con2}) for the stability of 
the uniform mixture. Finally, we comment that Eq. (\ref{con6})
can also be derived in a straightforward fashion from the  modulational 
instability analysis of an
appropriate set of dynamical equations, with $g_b|\psi_b|^2$ replaced by 
$\pi^2|\psi_b|^4$ in Eq. (\ref{m1}) and $\kappa=1/{4}$ in 
Eq. (\ref{m2}).

\section{Mixture with open boundary conditions} 

In this section we consider the 1D mixture with open boundary 
conditions. These boundary conditions can be used to model
a quasi-1D cylinder. In practice, if the axial length $L$ 
of the atomic wave-guide is much larger than the transverse width, 
i.e. $L \gg a_{\bot}j$, $j=b,f$, 
then one has a quasi-1D cylinder. The quasi-1D cylinder 
becomes infinite for $L\to \infty$. It this case we can 
use the Lagrangian density (\ref{l-tot}) with 
$z\in (-\infty, +\infty)$. With a finite number of particles 
(bosons and fermions) the uniform mixture in a infinite cylinder 
cannot exist, but localized solutions are instead possible 
with an attractive Bose-Fermi strength ($g_{bf}<0$) 
\cite{sala-st}. 

\subsection{Variational Results}

Here we develop a variational localized solution to Eqs. 
(\ref{mu1}) and (\ref{mu2}) 
noting that these equations can be derived from the Lagrangian 
\cite{VA} 
\beqa
\label{lag}
L&=&\int_{-\infty}^\infty\biggr[ \mu_b\phi_b^2
+\mu_f\phi_f^2-(\phi_b') ^2- (\phi_f')^2
-\phi_b^6 G(g_b/2\phi_b^2)
\nonumber \\ 
&-& \phi_f^6 F(g_f/2\phi_f^2)-g_{bf}\phi_b^2\phi_f^2\biggr]
dz -\mu_bN_b - \mu_fN_f
\eeqa
by demanding $\delta L/\delta \phi_b=\delta L/\delta 
\phi_f=\delta L/\delta \mu_b=\delta L/\delta \mu_f=0$.

To develop the variational approximation we use the following 
Gaussian ansatz \cite{sala-variational} 
\beqa
\phi_b(z)&=&\pi ^{-1/4}\sqrt{\frac{N_b A_b}{w_b}}
\exp \left(
-\frac{z^{2}}{2w_b^{2}}\right) ,\\
\phi_f(z)&=&\pi ^{-1/4}\sqrt{\frac{N_f A_f}{w_f}}
\exp \left(
-\frac{z^{2}}{2w_f^{2}}\right) ,
\label{Gauss}
\eeqa
where the variational parameters are $A_j$, 
the solitons' norm, and $w_j$ 
widths, in addition to $\mu_j$. Note that 
this Gaussian ansatz can be extended to include time dependence 
\cite{sala-variational-time}, as done recently to investigate 
the collective oscillation of a quasi-1D mixture made 
of condensed bosons and normal fermions \cite{sala-st}. 

The substitution of this variational ansatz in Lagrangian 
(\ref{lag}) yields 
\beqa
L&=&\mu_b N_b
A_b+\mu_fN_fA_f-\frac{N_bA_b}{2w_b^{2}}-
\frac{N_fA_f}{2w_f^{2}}
\nonumber \\
&-&\frac{A_b^3N_b^3}{\pi \sqrt 3 
w_b^2}G\left( \frac{cg_bw_b}{2N_bA_b}\right)
-\frac{A_f^3N_f^3}{\sqrt 3 \pi w_f^2}
F\left( \frac{dg_fw_f}{2N_fA_f}\right)
\nonumber \\
&-&\frac{g_{bf}N_bN_fA_bA_f}{\sqrt{\pi(w_f^2+w_b^2)}}
-\mu_bN_b-\mu_fN_f ,  \label{LG}
\eeqa
with $c=\sqrt{3\pi/2}$. The integrals over the $G(x)$ and $F(x)$ 
functions cannot be evaluated exactly for all $x$. However, the term 
involving the LL $G(x)$ function can be evaluated analytically
in the GP ($x\to 0$) and the TG  ($x\gg 1$) 
limits. The analytic term involving the $G(x)$ function 
in Eq. (\ref{LG}) 
is exact in both the limits  and provides a good description of the 
integral at other values of argument provided we take for the 
constant 
$c=\sqrt{3\pi/2}$.
The fermionic integral over the GY function 
$F(x)$ is evaluated similarly. There is no obvious   reason for choosing 
the constant $d$ in this integral. After a little experimentation 
it is taken 
to be $d=c= \sqrt{3\pi/2}$ which provides a faithful analytical  
representation of the numerical result. 

The first variational equations emerging from Eq.
(\ref{LG}) 
$\partial L/\partial \mu_b=\partial L/\partial \mu_f=0$
yield $A_b=A_f=1$. Therefore the conditions $A_b=A_f=1$
will be substituted in the subsequent variational 
equations. The variational equations 
$\partial L/\partial 
w_j=0$ lead to 
\beqa
&1&+ \frac{2N_b^2}{\pi\sqrt 3}G\left( 
\frac{cg_bw_b}{2N_b}\right)
-\frac{N_bg_b w_b}{2\sqrt{2\pi}}
G'\left( \frac{cg_bw_b}{2N_b}\right)
\nonumber \\
&+&\frac{g_{bf}N_fw_b^4}{\sqrt\pi  
(w_b^2+w_f^2)^{3/2}}=0,  \label{WG1}\\
&1&+\frac{2N_f^2}{\sqrt 3 \pi}F\left( \frac{cg_fw_f}{2N_f}\right)
-\frac{N_fg_f w_f}{2\sqrt{2\pi}}
F'\left( \frac{cg_fw_f}{2N_f}\right)\nonumber \\
&+&\frac{g_{bf}N_bw_f^4}{\sqrt\pi  
(w_b^2+w_f^2)^{3/2}}
=0.  \label{WG2}
\eeqa

The remaining variational equations  are
$\partial L/\partial A_j=0$, which
yield $\mu $ as a function of $w_j$'s,  and $g$'s:
\beqa
\mu_b &=& \frac{1}{2w_b^{2}}
-\frac{N_bg_b}{2 w_b\sqrt{2\pi}}
G'\left( \frac{cg_bw_b}{2N_b}\right)
\nonumber \\
&+&
\frac{\sqrt 3N_b^2}{\pi w_b^2}
G\left( \frac{cg_bw_b}{2N_b}\right)
+\frac{g_{bf}N_f}{\sqrt{\pi(w_f^2+w_b^2)}}.  \label{muG1}\\
\mu_f 
&=& \frac{1}{2w_f^{2}}+
\frac{\sqrt 3N_f^2}{\pi w_f^2}
F\left( \frac{cg_fw_f}{2N_f}\right)\nonumber \\
&-&\frac{N_fg_f}{2 w_f\sqrt{2\pi}}
F'\left( \frac{cg_fw_f}{2N_f}\right)
+\frac{g_{bf}N_b}{\sqrt{\pi(w_f^2+w_b^2)}}.  
\label{muG2}
\eeqa
Equations (\ref{WG1}) $-$ (\ref{muG2}) are the variational results which 
we shall use in our study of bright Bose-Fermi solitons. 

\begin{figure}[tbp]
\begin{center}
{\includegraphics[width=.8\linewidth]{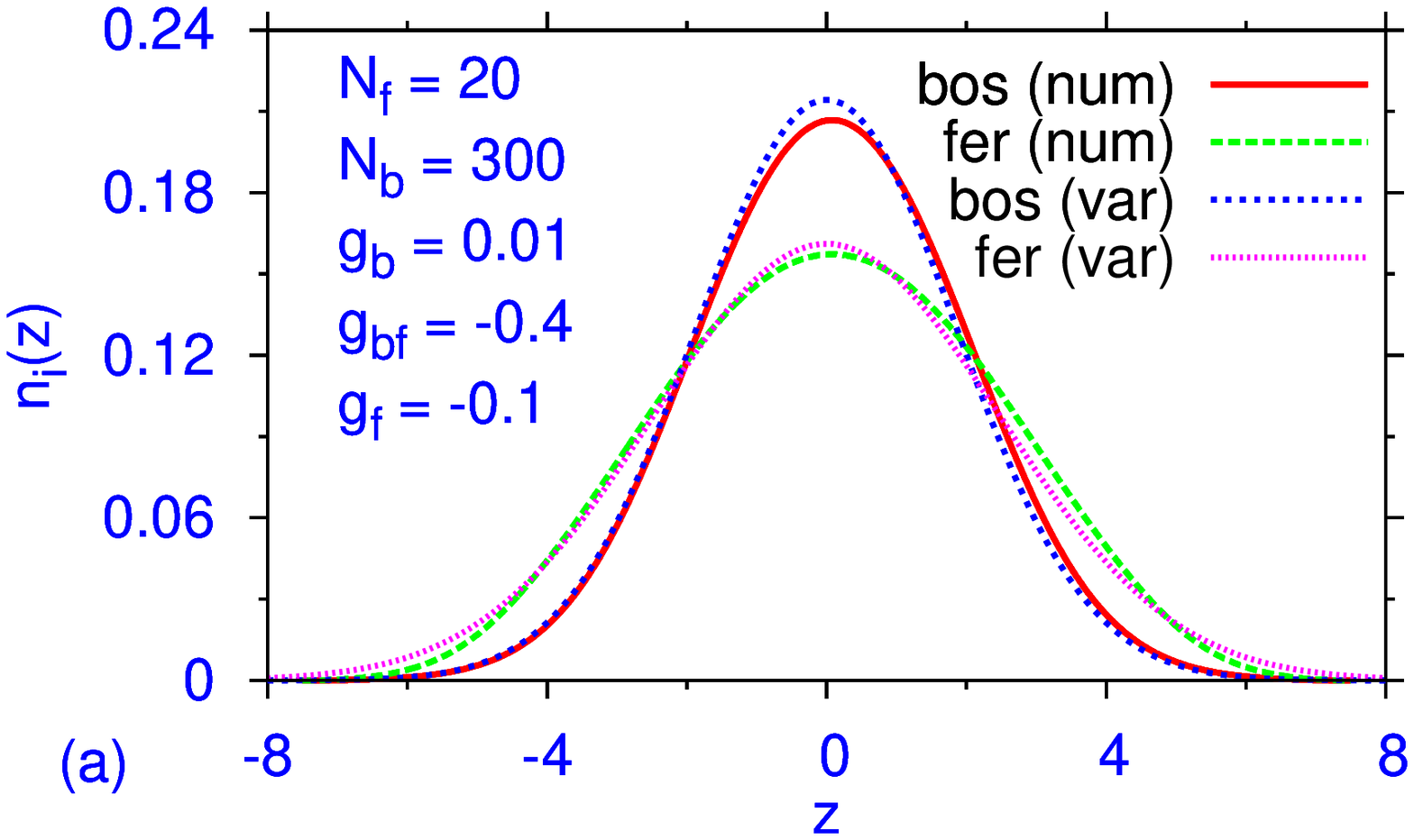}}
{\includegraphics[width=.8\linewidth]{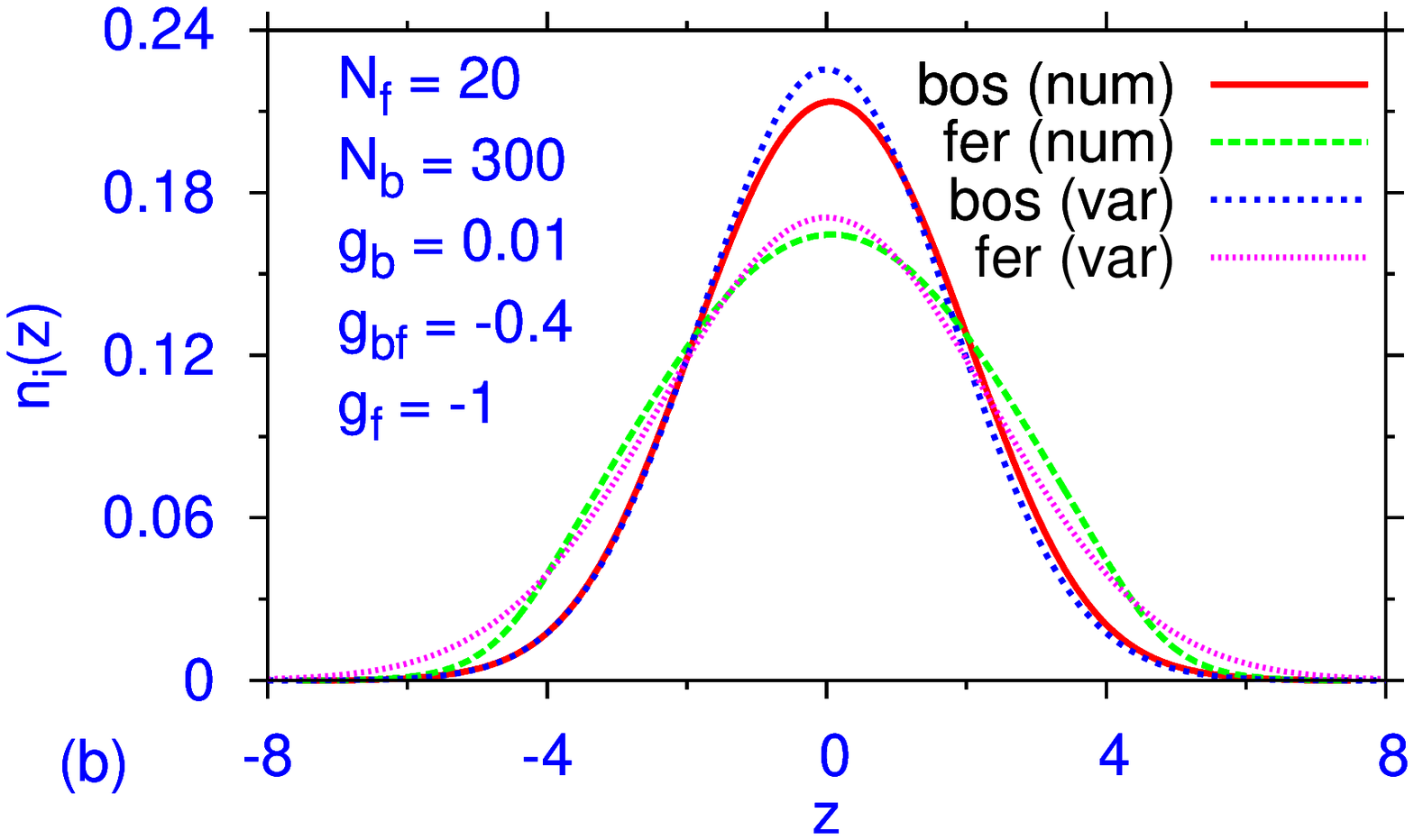}}
{\includegraphics[width=.8\linewidth]{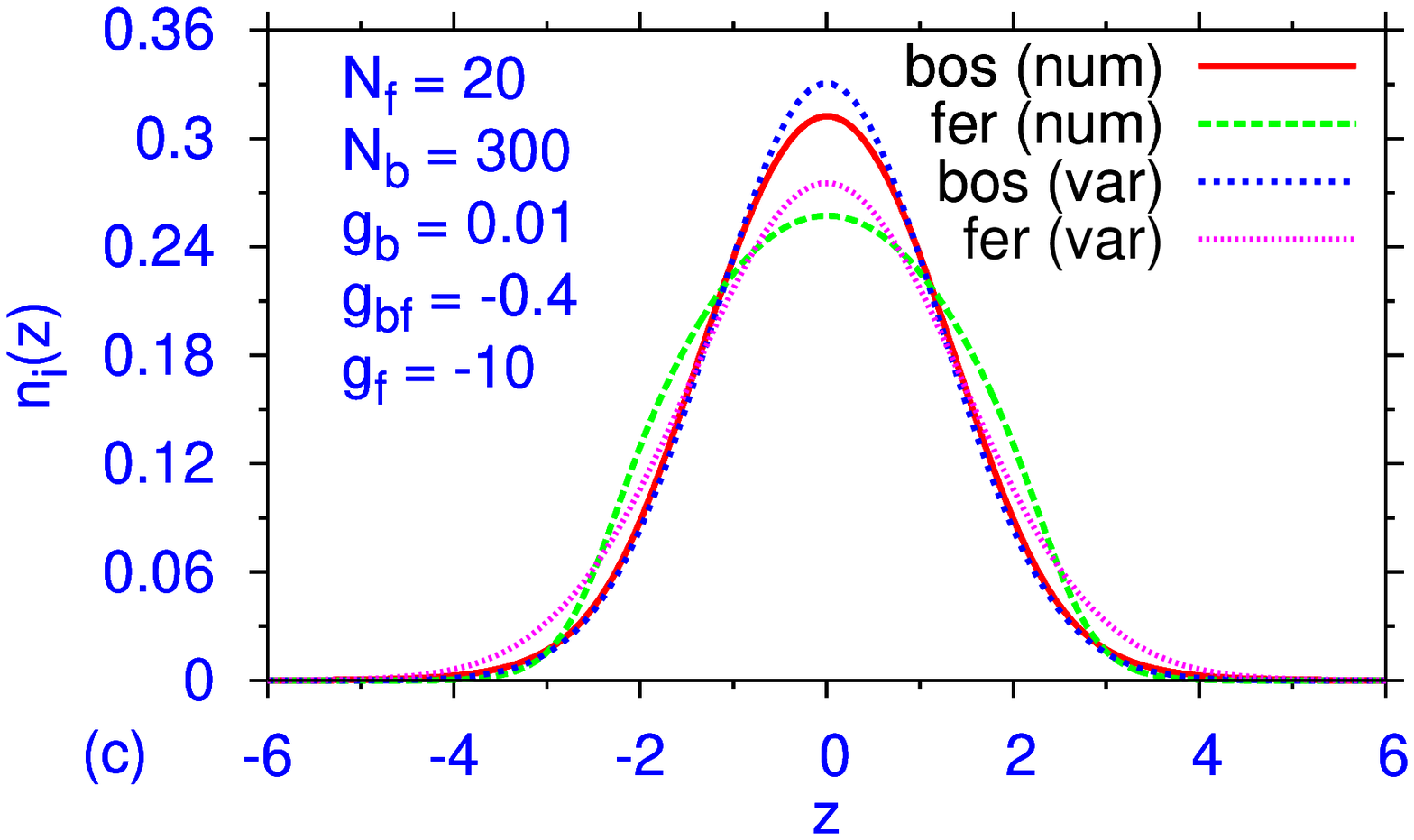}}
{\includegraphics[width=.8\linewidth]{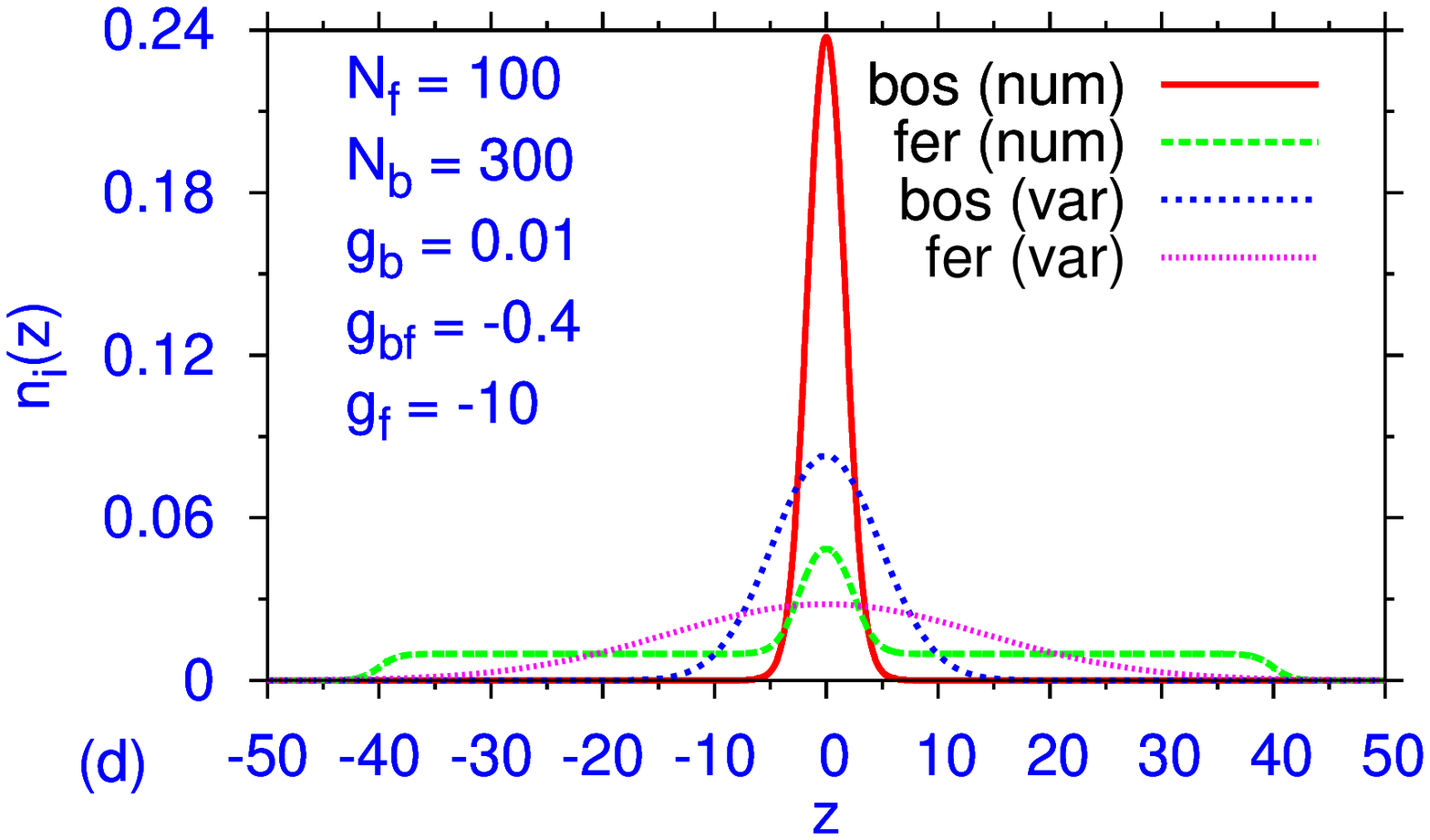}}
\end{center}
\caption{(Color online)
Probability densities from numerical 
solution of Eqs. (\ref{str1}) and (\ref{str2})  (here normalized to 
unity: $\int_{-\infty}^\infty n_i(z) dz=1 $)
compared with
variational results given by Eqs.  (\ref{WG1}) and (\ref{WG2}) for
$N_b=300, g_b=0.01, g_{bf}={-0.4}$ and (a) 
$N_f={20}, 
g_f=-0.1$, (b) 
$N_f=15, g_f=-1$, (c)  $N_f={20}, g_f=-10$, and (d)  
$N_f={100}, g_f=-10$. 
Of these, (a) represents fermions in BCS regime,  
(b) represents fermions in the BCS-to-unitarity crossover, 
(c) represents fermions in unitarity regime, 
and (d) represents bosons and fermions in a partially 
demixed configuration. } 
\label{fig2}
\end{figure}

\subsection{Numerical Results}

For stationary solutions we solve time-independent 
Eqs. (\ref{mu1}) and (\ref{mu2}) by using an 
imaginary time propagation method based on the finite-difference 
Crank-Nicholson discretization scheme of time-dependent 
Eqs. (\ref{str1}) and (\ref{str2}). The non-equilibrium 
dynamics from an initial stationary state
is studied by solving the time-dependent Eqs. (\ref{str1}) and
(\ref{str2}) with real time propagation by using as initial input the  
solution obtained by the imaginary time propagation method. The reason 
for this mixed treatment is that the imaginary time propagation method 
deals with real variables only and provides very accurate 
solution of the stationary problem at low computational 
cost \cite{muru}. In the finite-difference discretization we use space 
step of 0.025 and time step of 0.0005. 

\begin{figure}[tbp]
\begin{center}
{\includegraphics[width=.8\linewidth]{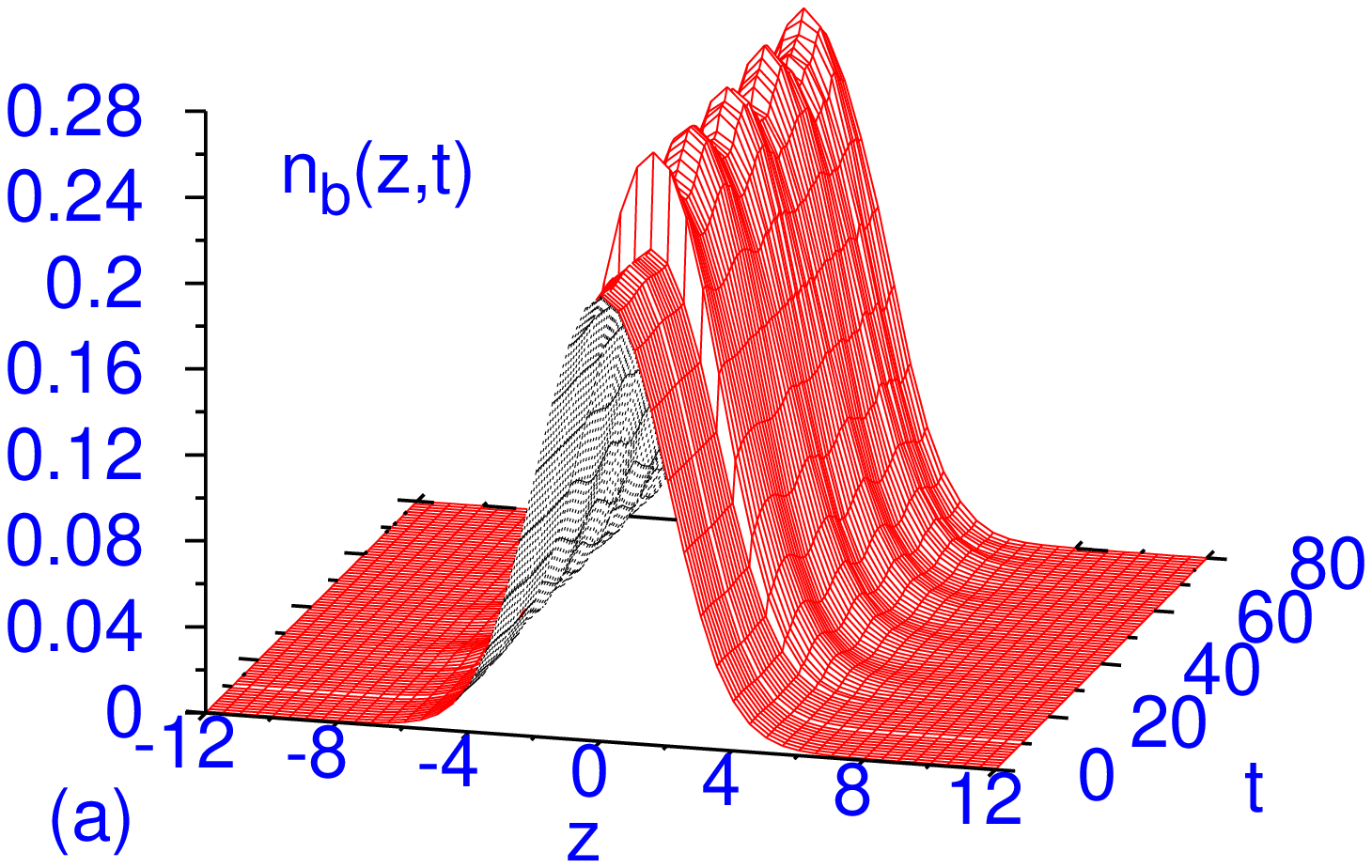}}
{\includegraphics[width=.8\linewidth]{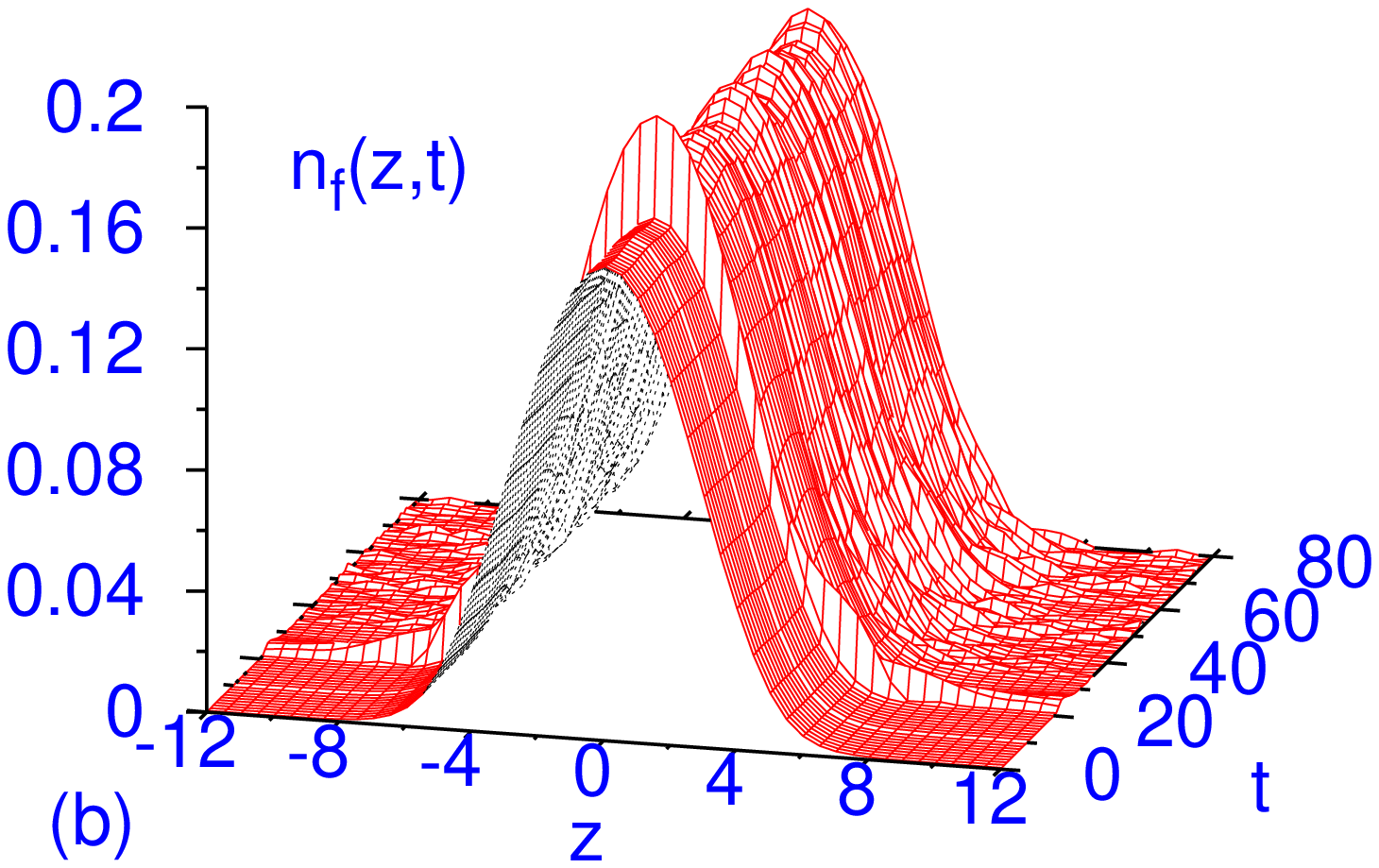}}
\end{center}
\caption{(Color online)
Dynamics of the probability density profiles of (a)  Bose and (b) Fermi 
solitons of Fig
\ref{fig2} (a) when at $t=20$ they are subject to a quite strong 
perturbation by setting $\phi_j(z,t) = 1.1\times \phi_j(z,t).$
The solitons undergo stable propagation as long as we could continue 
numerical simulation. The initial soliton profile is calculated with 
imaginary time propagation algorithm and the dynamics studied with 
real time propagation algorithm. The soliton profiles are normalized to 
unity: $\int_{-\infty}^\infty n_j(z,t) dz =1$. 
}
\label{fig3}
\end{figure}

First we report results for stationary profiles of the localized 
Bose and Fermi solitons formed in the presence of attractive Bose-Fermi 
and Fermi-Fermi interactions and repulsive Bose-Bose interactions. In 
the presence of weak attractive Fermi-Fermi interactions, the fermions 
form a BCS state which satisfy a nonlinear Schr\"odinger equation 
with repulsive (self defocussing) quintic nonlinearity. As the strength 
of the attractive Fermi-Fermi interaction increases the fermions pass 
from the BCS regime to the unitarity regime which is described by 
another nonlinear Schr\"odinger equation 
with repulsive (self defocussing) nonlinearity. 
Hence fermions cannot form a bright soliton by itself. 
However, they can form a bright soliton in the presence of an
attractive Bose-Fermi interaction.  

If the fermionic repulsive nonlinearity is not very large, bosons and 
fermions form mostly overlapping (mixed) solitons both in the BCS and 
unitarity regime. Note that in the BCS and unitarity regime 
the fermionic system becomes repulsive in the presence 
of Fermi-Fermi attraction. However, 
as the fermionic repulsive nonlinearity turns large the fermionic 
profile comes out of the bosonic profile and partially demixed solitons are 
created. We studied the numerically calculated soliton profiles 
for wide range of Bose-Bose, Bose-Fermi, 
and Fermi-Fermi interactions and boson and fermion numbers 
solving Eqs. (\ref{mu1}) and
(\ref{mu2}) by the technique of imaginary time propagation 
and compared them with the Gaussian variational results obtained from 
Eqs. (\ref{WG1}) and (\ref{WG2}). Except in the case of strong demixing,  
when the fermion profile strongly deviates from the Gaussian shape, the 
agreement between variational and numerical profiles is quite good.

\begin{figure}[tbp]
\begin{center}
{\includegraphics[width=.8\linewidth]{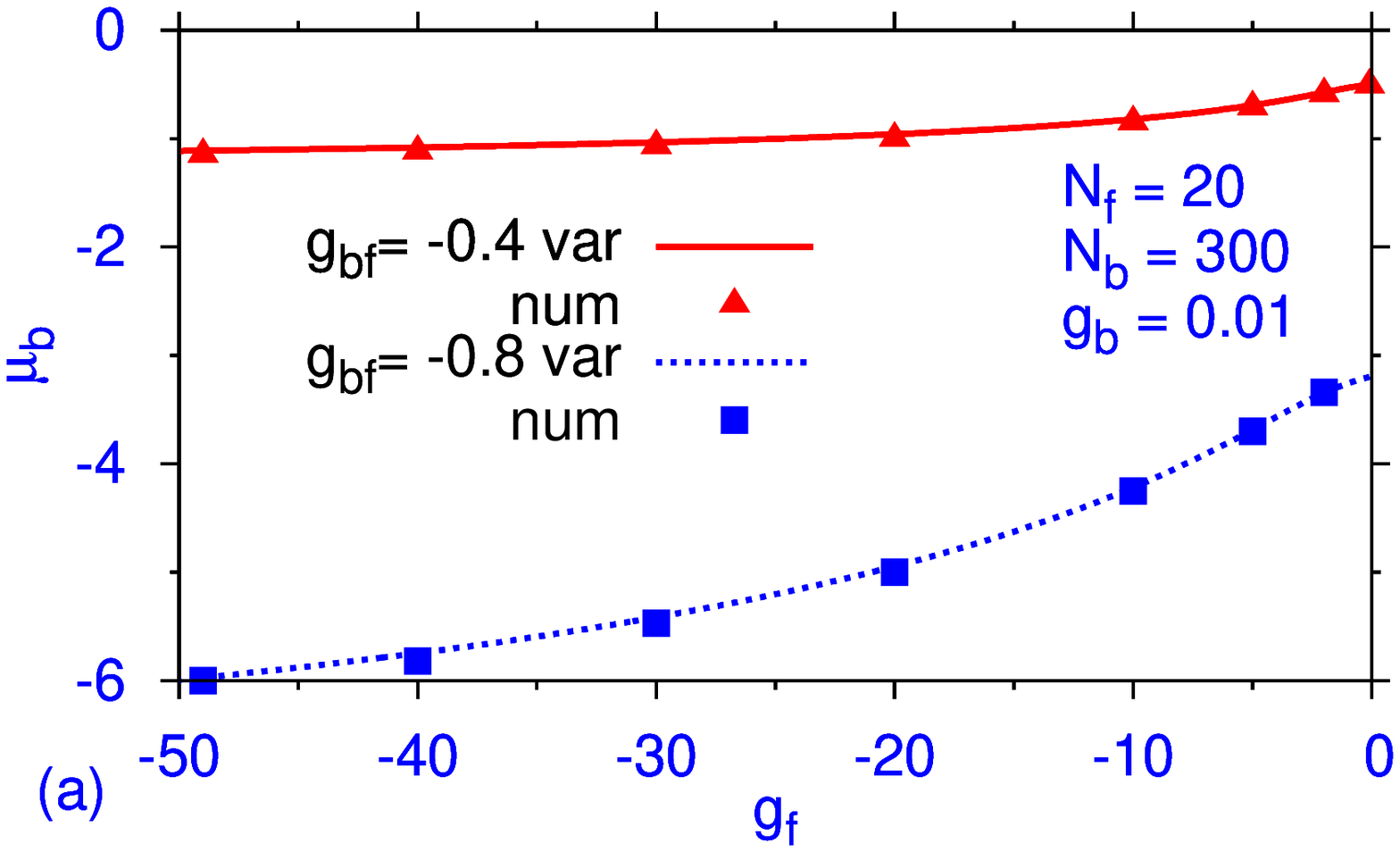}}
{\includegraphics[width=.8\linewidth]{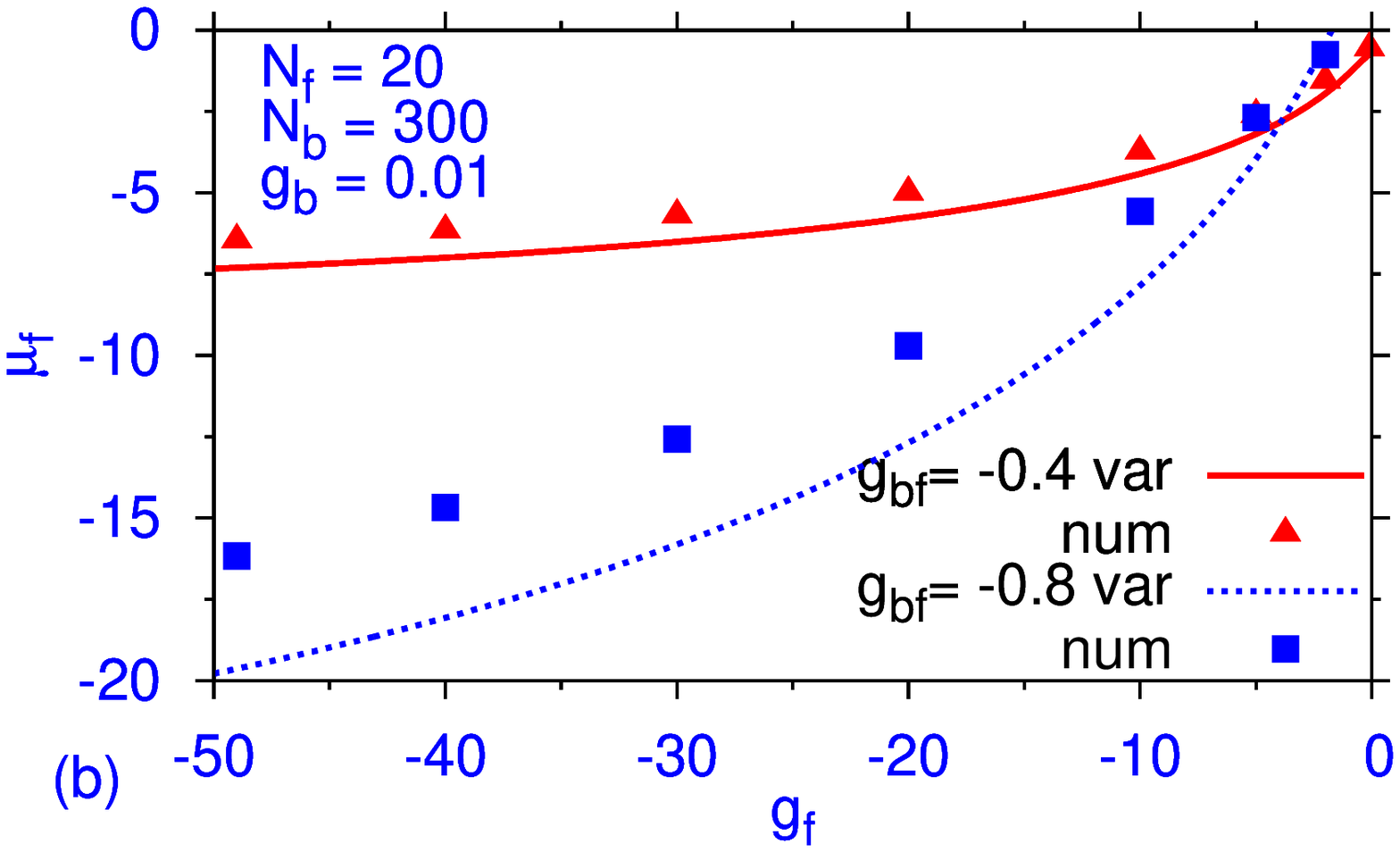}}
\end{center}
\caption{(Color online)
Chemical potentials for (a) bosons and (b) fermions vs. Fermi-Fermi 
interaction strength $g_f$ for $N_f={20}$, $N_b=300$, 
$g_b=0.01$ for 
$g_{bf} = -{0.4}$ and $-{0.8}$  calculated 
numerically (labelled ``num" shown 
by symbols)
compared with variational results (labelled ``var" shown by continuous 
lines). }
\label{fig5}
\end{figure}

In Fig. \ref{fig2} we present typical soliton profiles 
illustrating the 
change in the results during BCS-unitarity crossover as well as the demixed 
profiles. In Fig.  \ref{fig2} (a), (b), and (c) we show the soliton 
profiles for weak (BCS phase), moderate, and strong (unitarity regime) 
Fermi-Fermi attraction 
corresponding to strengths $g_f =-0.1, -1$ and $-10$, respectively, for 
$g_b=0.01$, $g_{bf}=-0.{4}, N_f={20},$ and 
$N_b=300$  and 
compare these with 
the variational results.  Figure   \ref{fig2}  (d) illustrates a 
demixed state 
obtained by increasing the fermion number from the configuration 
of Fig. \ref{fig2} (c) from $N_f={20}$ to 
{100}. In this case the (numerically 
obtained) fermion profile stretches far beyond the bosonic profile and 
is poorly represented by a Gaussian shape, which is the cause of 
deviation of the variational result from the numerical result.

After illustrating the soliton profiles in different states it is now 
pertinent to verify if these solitons are dynamically stable under 
perturbation. To this end we consider the typical stationary soliton of 
Fig. \ref{fig2} 
(a) (obtained by the imaginary time propagation method) and inflict on 
it the perturbation $\phi_j(z,t) = 1.1 \times \phi_j(z,t)$  and observe 
the resultant dynamics (obtained by the real time propagation 
method), which is illustrated in Fig. \ref{fig3}. The solitons under 
this perturbation execute breathing oscillation and propagate for as 
long as the numerical simulation was continued without being destroyed. 
This demonstrates the stability of the solitons under perturbation. 

Finally, in Figs. \ref{fig5} we show the chemical potential for 
$\mu_b$ bosons and $\mu_f$ for fermions 
as function of Fermi-Fermi interaction strength 
$g_f$ for $N_f={20}, N_b=300$, $g_b=0.01$, and 
$g_{bf}=-{.4}$  and ${-0.8}$ 
obtained from numerical solution and Gaussian variational analysis. The 
agreement between the two results for $\mu_b$ is good whereas for  
$\mu_f$ is only fair. The increase of the Fermi-Fermi attraction 
strength $|g_f|$ for a fixed $g_{bf}$
corresponds to a reduction in both chemical potentials
signaling  strongly bound solitons.  The same happens for the increase 
of Bose-Fermi attraction strength $|g_{bf}|$ from {0.4 
to 0.8}. In Figs. 
\ref{fig5} the small $|g_f|$ limit corresponds to the BCS phase of bosons 
whereas the large $|g_f|$ limit corresponds to the unitarity regime of 
fermions. The intermediate values of $|g_f|$ denote 
the crossover from BCS to the unitarity regime.

\section{Summary and Conclusion}

In this paper we have studied a one-dimensional superfluid Bose-Fermi 
mixture using a 
set of coupled mean-field equations derived as the Euler-Lagrange 
equation employing the Lieb-Liniger energy density of bosons and 
the Gaudin-Yang energy density of fermions and an interaction term 
proportional to the product of probability density of bosons and 
fermions. This set of coupled equations has a complex nonlinearity 
structure for fermions and bosons and 
shows the proper transition from a cubic Bose nonlinearity  
in the weak-coupling GP  BCS limit to a quintic nonlinearity in the 
strong-coupling TG  (Tonks-Girardeau) limit. In addition, for  
fermions with attractive interactions considered in this paper, 
it shows the proper transition from the weak-coupling BCS regime 
to unitarity limit: both limits 
are described by a quintic nonlinearity with different coefficients. 
In this model, in the extreme weak-coupling BCS limit the superfluid 
fermionic energy density is identical to that of a non interacting 
degenerate Fermi gas in the normal state \cite{sala-st}. 

We consider two distinct situations for the superfluid Bose-Fermi 
mixture: (i) in a ring with periodic boundary condition realizable 
from a toroidal trap in the limit of strong transverse confinement and 
(ii) in an infinite cylinder with open boundary condition 
realizable from an axially-symmetric 
trap in the limit of strong transverse and zero axial 
confinement. 

For the mixture in a ring, from energetic considerations, we obtain the
condition of stability of a uniform mixture with a constant probability
density. The uniform mixture is energetically stable for interspecies
attraction strength $|g_{bf}|$ below a critical value, above which
stable lowest-energy states are bright Bose-Fermi solitons. For
repulsive interspecies interaction the stable lowest-energy states are
demixed states of Bose-Fermi mixture, where the region of a maximum of
boson probability density corresponds to a minimum of fermion
probability density. It is also demonstrated algebraically that for 
attractive  Bose-Fermi interaction the bright solitons can be created 
from the uniform mixture above the critical Bose-Fermi attraction by 
modulational instability of the  uniform mixture under a 
weak perturbation.   

In the one-dimensional infinite cylinder we solved the coupled set of 
equations for the superfluid Bose-Fermi mixture numerically and using a 
Gaussian variational approximation. We calculated numerically the  
probability density profiles of the bright solitons as well as their  
chemical potentials  and compared them with the respective Gaussian 
variational approximations. The agreement between the two is good to 
fair. In this case a partial demixing of the Bose-Fermi solitons is 
possible, when the Fermi soliton extends over a very large region in 
space  while the Bose soliton remains fairly localized. We also 
established numerically the dynamical stability of the Bose-Fermi 
solitons by inflicting a perturbation on the solitons by multiplying the 
wave-function profiles by 1.1. The system is then found to propagate 
over a very long period of time executing breathing oscillation without 
being destroyed, which demonstrated the stability of the solitons. 
Finally, we comment that in view of the experimental  realization of the 
superfluid Bose-Fermi mixture \cite{bcsexp} and observation of solitons 
in a pure BEC \cite{Li-soliton,Rb-soliton},
the achievement of a bright Bose-Fermi soliton seems possible through a 
controlled manipulation of strengths of atomic interactions by 
varying an external background magnetic field near a 
Feshbach resonance \cite{fesh} and by adopting 
the set of parameters we have used in this paper. 

S.K.A. thanks  FAPESP and CNPq for partial financial support. 
L.S. acknowledges partial financial support by the Italian GNFM-INdAM 
through the project ``Giovani Ricercatori'' and by   Fondazione 
CARIPARO.


\begin{thebibliography}{99}
\bibitem{pade-b} The Lieb-Liniger function $G(x)$ is such that
for $x\ll 1$ one finds $G(x)\simeq x+Bx^{3/2}+\eta x^2$,
where $B=4/(3\pi)$ and $\eta=0.0648$ (from numerics).
Instead for $x\gg 1$ one has $G(x)\simeq (\pi^2/2)x^2/(x+2)^2$. 
$G(x)$ can be very well described by the following Pad\`e approximant:  
$G(x)=(x+Ax^3)/[1+Bx^{1/2}+Cx(x+2)^2]$, where $A=\pi^2(B^2-\eta )/12$
and $C=(B^2- \eta)/4$.
\bibitem{pade-f} 
The Gaudin-Yang function $F(x)$ is conveniently parametrized 
by the following Pad\`e approximant:  
$F(x) = (\pi^2/12)(x^2 - x + 3/4)/(x^2 + q_1 x + q_2)$, 
where $q_1 = - [ 9/(8 \pi^2 ) + 1/4] = - 0.3633$ and $q_2 = 3/16$.
\end{thebibliography}
\end{document}